\documentstyle[12pt]{article}
\renewcommand{\O}{\Omega}
\renewcommand{\o}{\omega}
\newcommand{\g}{{\scriptstyle \cal G\mit}}
\newcommand{\f}{\frac{k}{4\pi}}
\newcommand{\F}{{\cal F}}
\renewcommand{\d}{\delta}
\newcommand{\m}{{\cal M}}
\newtheorem{theorem}{Theorem}
\newtheorem{lemma}{Lemma}
\newtheorem{definition}{Definition}

\begin{document}
\begin{titlepage}
\title{\begin{flushright} {\normalsize UUITP 32/1993   \\
hep-th/9312004 } \\
\end{flushright}  \bigskip   \bigskip \bigskip   \bigskip  \bigskip
{\bf Symplectic structure
of the moduli space\\ of flat connection on a Riemann surface}}
\author{\\ \\ {\bf A.Yu. Alekseev}
\thanks{On leave of absence from Steklov Mathematical Institute,
Fontanka 27, St.Petersburg, Russia}
\thanks{Supported by Swedish Natural Science Research Council
(NFR) under the contract F-FU 06821-304}
\\
{\bf A. Z. Malkin}
\thanks{On leave of
absence from St.-Petersburg University.}
\thanks{Supported
in part by a Soros Foundation Grant awarded by
the American Physical Society.}
\\\\
Institute of Theoretical Physics, Uppsala University,
\\ Box 803 S-75108, Uppsala, Sweden.\\\\}
\date{November 1993}
\maketitle\thispagestyle{empty}
\begin{abstract}
We consider canonical symplectic structure on the moduli
space of flat  ${\g}$-connections on a Riemann surface
of genus $g$ with $n$ marked points. For ${\g}$ being
a semisimple Lie algebra we obtain an explicit efficient formula
for this symplectic form and prove that it may be represented
as a sum of $n$ copies of Kirillov symplectic form on the orbit
of  dressing transformations in the Poisson-Lie group $G^{*}$
and $g$ copies of  the symplectic structure on the Heisenberg
double of the Poisson-Lie group $G$ (the pair ($G,G^{*}$) corresponds
to the Lie algebra ${\g}$).
\end{abstract}
\end{titlepage}

\section{Introduction}
\setcounter{equation}{0}

Being  interesting object of investigations, the moduli space of flat
connections
on a Riemann surface attracted  attention of many physicists and
mathematicians
when its relation to the Chern-Simons theory had been discovered
\cite{Wt}.
By definition the moduli space (we shall often refer to the moduli
space of flat
connections in this way) is a quotient of the infinite dimensional
space of flat
connections over the infinite dimensional gauge group.  It is
remarkable that
this quotient appears to be finite dimensional.

The moduli space $\m$ carries a nondegenerate symplectic structure
\cite{AB}.
It implies the existence of a nondegenerate Poisson bracket on $\m$.
Recently the combinatorial description of the moduli space has been
suggested \cite{Fock}. The main idea is to represent the same space
$\m$ as
a quotient of the finite dimensional  space ${\cal P}$ over the
finite dimensional
group action. The Poisson structure has been defined on  ${\cal P}$
and proved to
reproduce the canonical Poisson structure on the moduli space after
reduction.

In the first part of this paper we give a combinatorial description
of the canonical
symplectic structure on $\m$ (see {\em Theorem 1}, Section 3). This
is a bit more
natural object to consider because the symplectic form may be
canonically
mapped from $\m$ to  ${\cal P}$    by means of the pull-back, whereas
the Poisson
bracket  may be defined on ${\cal P}$ in many ways.

The nonabelian 3-dimensional Chern-Simons theory has been solved
because
it is related to 2-dimensional Wess-Zumino model and to the Quantum
Groups.
In particular, let us consider the Hilbert space ${\cal H}$ of the CS
theory associated to simple Lie algebra $\g$ on an equal time Riemann
surface $\Sigma$
of genus $g$ with $n$ marked points. By construction, there is a
representation $I_{i}$
assigned to each marked point. Then the Hilbert space  ${\cal H}$  is
isomorphic
to the space of invariants

\begin{equation}
{\cal H}=Inv_{q}(I_{1}\otimes \dots \otimes I_{n}\otimes \Re^{\otimes
g}) \label{I1}
\end{equation}
in the tensor product of the corresponding representations of the
quantum group
$U_{q}(\g)$. In formula (\ref{I1}), we denote by $\Re$ the regular
representation of
$U_{q}(\g)$ corresponding to a handle. In this paper we prove a
quasi-classical
analogue of this statement (see {\em Theorem 2}, Section 4).

The first attempt in this direction had been made in \cite{EMSS}.
There the cases of torus and a disc with one marked point had been
considered. However, the key object which will enter into the answer
appeared quite recently \cite{Gav}, \cite{Antons}. This is the set of
symplectic forms associated to Poisson-Lie groups which replace
quantum groups in the quasi-classical limit. More precisely, there is
a family
of symplectic forms $\vartheta$ on the orbits of dressing
transformations \cite{Sem}.
They are naturally assigned to the marked points. Besides we have a
symplectic
form $\theta$ on the so-called Heisenberg double (analogue of the
cotangent bundle)
which is responsible for a handle. So, we prove that the symplectic
structure on the moduli space of flat connections on a Riemann
surface may
be represented as a direct sum of $n$ copies of $\vartheta$ and $g$
copies of
$\theta$:
\begin{equation}
\O=\sum_{i=1}^{n}{\vartheta_{i}}+\sum_{i=1}^{g}{\theta_{i}}.
\label{I2}
\end{equation}

\section{Preliminaries}
\setcounter{equation}{0}

This section includes a collection of facts which we shall use
throughout the paper.

\subsection{Definition of the symplectic structure on the moduli
space}
Let  $\Sigma$ be a Riemann surface of genus $g$ with $n$ marked
points.
Consider a connection $A$ on $\Sigma$ taking values in a simple
Lie algebra ${\g}$. We denote the Killing form on ${\g} $ by $Tr$.
There
is a canonical symplectic structure on the space ${\cal A}$ of all
smooth
connections \cite{AB}:

\begin{equation}
\Omega_{\cal A}=\frac{k}{4\pi} Tr \int_{\Sigma}{\d A\wedge \d A}.
\label{SA}
\end{equation}
Here we have introduced a  coefficient $\frac{k}{4\pi}$
in order to make our notations closer to the ones
accepted in the physical literature.

The form (\ref{SA}) is obviously nondegenerate and invariant with
respect to the action of the gauge group $G_{\Sigma}$:

\begin{equation}
A^{g} = g^{-1} A g +g^{-1}dg. \label{Gtr}
\end{equation}
We denote the exterior derivative on the Riemann surface by $d$,
whereas the exterior derivative on the space of connections, moduli
space
or elsewhere is always $\d$.  The action (\ref{Gtr}) is actually
Hamiltonian and
the corresponding momentum mapping is given (up to a multiplier) by
the curvature:

\begin{eqnarray}
\mu(A)=-\frac{k}{2\pi} F;\nonumber \\
F=dA - A^{2}.   \label{F}
\end{eqnarray}

Let us start with a case when there is no marked points.
\begin{definition}
The space of flat connections $\Im_{g}$ on a Riemann surface of genus
$g$
is defined as a zero level surface
of the momentum mapping (\ref{F}):

\begin{equation}
F(z)=0.  \label{F0}
\end{equation}
\end{definition}

\begin{definition}
The moduli space of flat connections is a quotient of the space of
flat connections $\Im_{g}$ over the gauge group action (\ref{Gtr}):

\begin{equation}
{\cal M}_{g} =\Im_{g} /G_{\Sigma}.   \label{mg}
\end{equation}
\end{definition}
The curvature being the momentum mapping for the gauge group,
the moduli space may be obtained by Hamiltonian reduction from
the space of smooth connections. General theory of Hamiltonian
reduction
\cite{Arn},\cite{Wein} ensures that  the moduli space carries
canonical nondegenerate symplectic structure induced from the
symplectic  structure (\ref{SA}) on ${\cal A}$.

Now  we turn to more sofisticated case of  the Riemann surface with
marked
points.  Among several possible approaches we choose the one which is
more convenient for the further consideration.

To each marked point  $z_{i}$ we assign a coadjoint orbit in the
space $\g^{*}$
dual to the Lie algebra $\g$. Having the nondegenerate Killing form
on $\g$,
we can actually identify $\g$ and $\g^{*}$. In this case the
coadjoint orbit may
be viewed just as a conjugancy class in $\g$. Using a matrix
realization of
the Lie algebra we get

\begin{equation}
T\in {\cal O}_{D} \Leftrightarrow T=v^{-1}Dv, \quad v\in G. \label{O}
\end{equation}
Here $D$ is any element of  $\g$ which belongs to the orbit ${\cal
O}$.
For example, we can choose it in such a way that it will be
represented by
a diagonal matrix.  Any coadjoint orbit ${\cal O}_{D}$ carries a
nondegenerate
symplectic form \cite{Kir} which is often called Kirillov form. Using
$v$ and
$D$ instead of $T$ one can represent Kirillov form as

\begin{equation}
\varpi_{D}= Tr D(\d vv^{-1})^{2}.    \label{SO}
\end{equation}
It is easy to check that formula (\ref{SO}) indeed defines the
nondegenerate
closed two-form on the orbit ${\cal O}_{D}$ invariant with respect to
conjugations.
It is worth mentioning that  $T$ is a momentum mapping for the group
action

\begin{equation}
T^{g}=g^{-1}Tg, \quad v^{g}=vg.   \label{Conj}
\end{equation}

\begin{definition}
A decorated Riemann surface with $n$ marked points is a Riemann
surface
and a set of coadjoint orbits ${\cal O}_{1},\dots ,{\cal O}_{n}$
assigned to the marked
points $z_{1},\dots ,z_{n}$.
\end{definition}
One can use the notion of decoration in order to describe possible
singularities
which may be developed by connections at marked points. Let us
introduce the local
coordinate $\phi_{i}$ in the small neighborhood of the marked point
$z_{i}$ so that

\begin{equation}
\oint_{S_{i}}{d\phi_{i}}=2\pi.  \label{angles}
\end{equation}
Here $S_{i}$ is a closed contour which surrounds the marked point.
Apparently,
the coordinate $\phi_{i}$ measures the angle in the neighborhood of
$z_{i}$.
On the surface with marked points we shall admit connections which
have singularities of the form

\begin{equation}
A(z)_{z\sim z_{i}}=A_{i}d(\frac{\phi_{i}}{2\pi}) + \tilde{A}(z),
\label{SingA}
\end{equation}
where $A_{i}$ are constant coefficients and  $\tilde{A}(z)$ is a
smooth connection. We call the coefficients $A_{i}$ singular parts of
$A$.

\begin{definition}
The space of connections ${\cal A}_{g,n}$ on a decorated Riemann
surface
with marked points is defined by the requirement that the singular
parts of the
connection belong to the coadjoint orbits assigned to the
corresponding
marked points:

\begin{equation}
\frac{2\pi}{k} A_{i}\in {\cal O}_{i}.  \label{AO}
\end{equation}
\end{definition}
It is remarkable that the symplectic structure (\ref{SA}) may be used
for
the space ${\cal A}_{g,n}$ as well. It is convenient to introduce one
more
symplectic space which is the direct product of  ${\cal A}_{g,n}$ and
its
collection of coadjoint orbits:

\begin{equation}
{\cal A}_{g,n}^{tot}={\cal A}_{g,n}\times {\cal O}_{1}\times \dots
\times {\cal O}_{n}.  \label{Atot}
\end{equation}
It carries the symplectic structure

\begin{equation}
\Omega_{\cal A}^{tot}=\Omega_{\cal A}+\sum_{i}^{n}{\varpi_{i}},
\label{SAtot}
\end{equation}
The action of the gauge group may
be defined on the space ${\cal A}_{g,n}^{tot}$ as follows:

\begin{eqnarray}
A^{g} = g^{-1} A g +g^{-1}dg:   \nonumber \\
T_{i}^{g}=g(z_{i})^{-1}T_{i}g(z_{i}), \quad v_{i}^{g}=v_{i}g(z_{i}).
 \label{MGtr}
\end{eqnarray}
As we see, the modified gauge transformations are combined from the
standard
gauge transformations (\ref{Gtr}) and orbit conjugations
(\ref{Conj}). The momentum
mapping for the gauge group action (\ref{MGtr}) looks very similar to
(\ref{F}):

\begin{equation}
\mu(z)=\sum_{i}^{n}{T_{i}\d (z-z_{i})}-\frac{k}{2\pi} F(z).
\label{MF}
\end{equation}
It is easy to see that the definition of  ${\cal A}_{g,n}$ ensures
that there is a lot
of solutions of the zero level conditions.

\begin{definition}
The space of flat connections on a decorated Riemann surface
$\Im_{g,n}$
is defined as a space of solutions of the following equation which
replaces
the zero curvature condition:

\begin{equation}
\mu(z)=0.     \label{MF0}
\end{equation}
\end{definition}

Let us choose a  loop $S_{i}$ surrounding the marked point $z_{i}$.
One can define the monodromy matrix (or parallel transport)  $M_{i}$
along this way.
It is easy to check that if $A$ and $\{T_{i}\}$ satisfy  (\ref{MF0}),
the monodromy matrix $M_{i}$ belongs to the conjugancy class
of the  exponent of $D_{i}$
\begin{equation}
M_{i} = u_{i}^{-1} exp ( \frac{2\pi}{k} D_{i}) u_{i}.
\label{CExporb}
\end{equation}

\begin{definition}
The moduli space of flat connections on a Riemann surface of genus
$g$ with
$n$ marked points $\m_{g,n}$ is defined as a quotient of the space of
flat connection on
a decorated Riemann surface over the gauge group action (\ref{MGtr}):

\begin{equation}
\m_{g,n} =\Im_{g,n} /G_{\Sigma}.  \label{Mgn}
\end{equation}
\end{definition}
It is important that the moduli space $\m_{g,n}$ is obtained by
Hamiltonian reduction
from the symplectic space ${\cal A}_{g,n}^{tot}$. This procedure
provides the
nondegenerate symplectic form on  $\m_{g,n}$ which is the main object
of this paper.

Let us finish this subsection by the remark that  symplectic spaces
$\m_{g,n}$
naturally appear as phase spaces  in the Hamiltonian Chern-Simons
theory
(see for example \cite{EMSS}). So, the results concerning the
corresponding
symplectic forms may be always reinterpreted on the language of the
Chern-Simons theory.

\subsection{Poisson-Lie groups}
Let us consider the Lie group $G$ associated to the Lie algebra $\g$.
We shall introduce a Poisson bracket on $G$ such that the
multiplication:
\begin{equation}
G\times G\rightarrow G
\end{equation}
is a Poisson mapping. A Lie group endowed with such a Poisson bracket
is called a Poisson-Lie group. To give an expression for this bracket
we need some notations. Let $r_{+}$ and $r_{-}$ be classical
$r$-matrices corresponding to the Lie algebra $\g$:
\begin{equation} r_{+}=\frac{1}{2}\sum h_{i}
\otimes h^{i}+\sum_{\alpha \in \Delta_{+}}e_{\alpha }
\otimes e_{-\alpha }, \ \   \end{equation}
\begin{equation} r_{-}=-\frac{1}{2}\sum h_{i}
\otimes h^{i}-\sum_{\alpha \in \Delta_{+}}e_{-\alpha }
\otimes e_{\alpha }.  \ \  \end{equation}
Then the Poisson bracket on the matrix elements of the group $G$ is
the following \cite{Skl}:
\begin{equation}
\{g^{1},g^{2}\}=[r_{+},g^{1}g^{2}]=[r_{-},g^{1}g^{2}].
\label{PBG}
\end{equation}
Here we use  tensor notation $g^{1}=g\otimes I$, $g^{2}=I\otimes g$.
A simple Lie group $G$ equipped with brackets (\ref{PBG}) is a
Poisson-Lie group.
Another  Poisson-Lie group which we need is called $G^{*}$.  An
element of $G^{*}$ is a pair $(L_{+},L_{-})$, where $L_{+}$ ($L_{-}$)
is an element of Borel subgroup generated by positive (negative)
roots of $\g$. The Cartan part of $L_{+}$ is inverse to the one of
$L_{-}$.
The multiplication on the group $G^{*}$ is component-wise:
\begin{equation}
(L_{+},L_{-})(L_{+}^{'},L_{-}^{'})=(L_{+}L_{+}^{'},L_{-}L_{-}^{'}).
\end{equation}
The Poisson bracket on $G^{*}$ looks like follows \cite{Sem}:
\begin{eqnarray}
\{L_{+}^{1},L_{+}^{2}\}=[r_{+},L_{+}^{1}L_{+}^{2}],  \nonumber \\
\{L_{-}^{1},L_{-}^{2}\}=[r_{+},L_{-}^{1}L_{-}^{2}],    \\
\{L_{+}^{1},L_{-}^{2}\}=[r_{+},L_{+}^{1}L_{-}^{2}].   \nonumber
\end{eqnarray}

It is useful to introduce  a mapping $\alpha$ from $G^{*}$ to $G$
\begin{equation}
\alpha: (L_{+}, L_{-}) \rightarrow L=L_{+} L_{-}^{-1}. \label{LL}
\end{equation}
 The groups structures of $G$ and $G^{*}$ are different and the
mapping $\alpha$
is not a group homomorphism. However,  we shall see in Section 4 that
it may be useful
if we replace the requirements of group homomorphism by some weaker
conditions.

The matrix elements of the resulting  element $L$  have the following
Poisson bracket:
\begin{equation}
\{ L^{1},L^{2}\} =r_{+}L^{1}L^{2}
+L^{1}L^{2}r_{-}-L^{1}r_{+}L^{2}-L^{2}r_{-}L^{1}.    \label{PBG*}
 \end{equation}

The Poisson bracket  (\ref{PBG*}) is degenerate. So, one can
describe its symplectic leaves.
To this end we consider the action of $G$ on $G^{*}$ by means of
dressing
transformations \cite{Sem}:
\begin{equation}
L\rightarrow g^{-1}Lg,   \quad   L\in G^{*}, \quad g\in G
\label{DT}
\end{equation}
This action is a Poisson one. It means that the mapping
\begin{equation}
G\times G^{*}\rightarrow G^{*}
\end{equation}
is consistent with Poisson structures on $G$ and $G^{*}$.

Dressing transformations are  useful  when one describes  symplectic
structures
associated to $G^{*}$.  The result has been obtained in two steps.
First, it was proved \cite{Sem} that symplectic leaves of the Poisson
bracket (\ref{PBG*}) are orbits of dressing transformations
(\ref{DT}) and then  the  expression for the symplectic forms was
found in   \cite{Gav}, \cite{Antons}. To write down the answer we
choose a particular orbit of dressing transformations:
\begin{equation}
L=g^{-1}Cg,     \quad   L\in G^{*}, \quad g\in G,    \label{Orb}
\end{equation}
where C is an element of Cartan subgroup which parametrizes the
orbit. So we have the mapping $\pi :G\rightarrow G^{*}$ given by
(\ref{Orb}).
It is convenient to use coordinates $L_{+}, L_{-}$ and $g$ on the
orbit  simultaneously. We have the following formula for the
pull-back of the symplectic form on the orbit (\ref{Orb}) along the
projection $\pi$:
\begin{equation}
\vartheta(g,C) =\frac{1}{2}Tr\{ C\d gg^{-1}
\wedge C^{-1}\d gg^{-1}+L_{+}^{-1}\d L_{+}
\wedge L_{-}^{-1}\d L_{-}\}
\label{FG}
\end{equation}
We shall see in Section 4 that  the orbit of dressing transformations
may be naturally
associated to each marked point on the Riemann surface.

Now we have a full analogue  of the classical theory of coadjoint
orbits of the group $G$  for the Poisson-Lie case. The dressing
transformations replace the coadjoint action and form  (\ref{FG})
replaces Kirillov form (\ref{O}). To complete the program we should
find an object which corresponds to the cotangent bundle $T^{*}G$.
Actually, it has been introduced in \cite{Sem} and  called Heisenberg
double $D_{+}$.  In the  case at hand (simple Lie group with Poisson
brackets (\ref{PBG})) $D_{+}$ is isomorphic to the Cartesian product
of two copies of $G$:
\begin{equation}
D_{+}\simeq G\times G
\end{equation}
So $D_{+}$ is a Lie group with component-wise multiplication. There
exists a Poisson structure on $D_{+}$ such that the following
embeddings of $G$ and $G^{*}$ into $D_{+}$ are Poisson mappings:
\begin{eqnarray}
G\rightarrow D_{+} :\quad g\rightarrow (g,g)   \\
G^{*}\rightarrow D_{+} :\quad L\rightarrow (L_{+},L_{-})
\label{emb}
\end{eqnarray}
We do not write  this Poisson bracket (see for example \cite{Sem}),
but make two remarks about it. First, $D_{+}$ is not a Poisson-Lie
group ({\em i.e}. this bracket is not consistent with
multiplication). Second, the Poisson structure on $D_{+}$ is
degenerate, but there is the symplectic leaf
\begin{equation}
{\cal L}=GG^{*}\cap G^{*}G \label{L}
\end{equation}
which is open and dense in $D_{+}$.  In formula (\ref{L}) $G$,
$G^{*}$ are embedded
into $D_{+}$ by means of the mappings  (\ref{emb}). To write down the
symplectic form on this leaf  let us consider a set

\begin{equation}
\aleph=\{((g,L),(L^{'},g^{'}))\in (G\times G^{*})\times (G^{*}\times
G): gL_{+}=L_{+}^{'}g^{'},gL_{-}=L_{-}^{'}g^{'}\}.
\end{equation}
One can define a natural projection:

\begin{equation}
\aleph \rightarrow {\cal L} : ((g,L),(L^{'},g^{'}))\rightarrow
(gL_{+}=L_{+}^{'}g^{'},gL_{-}=L_{-}^{'}g^{'})\end{equation}
The pull-back of the symplectic form on ${\cal L}$ along this
projection is the following \cite{Antons}:

\begin{eqnarray}\label{FD}
\theta(g,g^{'},C) =\frac{1}{2}Tr\{ C\d gg^{-1}
\wedge C^{-1}\d gg^{-1}+L_{+}^{-1}\d L_{+}
\wedge L_{-}^{-1}\d L_{-}\}+   \nonumber \\
+\frac{1}{2}Tr\{ C^{-1}\d g^{'}g^{'-1}
\wedge C\d g^{'}g^{'-1}+L_{+}^{'-1}\d L^{'}_{+}
\wedge L_{-}^{'-1}\d L^{'}_{-}\}+ \\
+Tr \d CC^{-1}\wedge (\d g g^{-1}-\d g^{'} g^{'-1}). \nonumber
\end{eqnarray}

As one can see the symplectic form on $D_{+}$ consists of two terms
similar to
the symplectic forms on the orbits (\ref{FG}). So we have two orbit
systems (their
dynamical variables are denoted by letters $(g,L)$ and
$(g^{'},L^{'})$)  which contain
points $C$ and $C^{-1}$. The last term in (\ref{FD}) is designed to
take into account
the fact that now $C$ is a dynamical variable as well. The form
(\ref{FD}) will appear
in Section 4. It will correspond to the contribution of one handle
into symplectic form
on the moduli space.

\subsection{Dual pairs}
One of  powerful tools in Hamiltonian mechanics is the language of
dual
pairs. Let $X$ be a symplectic space.  Obviously, it carries a
nondegenerate
Poisson structures.

\begin{definition}
A pair of Poisson mappings

\begin{eqnarray}
\mu : X\rightarrow Y, \nonumber \\
\nu :X\rightarrow Z      \label{DP}
\end{eqnarray}
is called a dual pair if

\begin{equation}
\{ \{f,h\}=0, \forall f= \tilde{f} \circ \mu, \tilde{f}:Y \rightarrow
C\} \Leftrightarrow
\{ \exists \tilde{h}:Z \rightarrow C, h= \tilde{h} \circ \nu\}.
\end{equation}
\end {definition}
In other words,  any function lifted from $Y$  is in involution with
any function
lifted from $Z$ and moreover, if some function commute with any
function lifted
from $Y$ it means that it is lifted from $Z$.

The standard source of dual pairs is Hamiltonian reduction. If we
have
a Hamiltonian action of a group $G$ on a symplectic manifold $X$, the
following pair of projections is dual:

\begin{eqnarray}
\mu :X\rightarrow \g^{*}, \nonumber \\
\nu :X\rightarrow X/G.   \label{HamR}
\end{eqnarray}
Here the mapping $\mu$ is the momentum mapping from the manifold $X$
to
the space dual to the Lie algebra $\g$.

Dual pairs provide the method to classify symplectic leaves in the
Poisson spaces
$Y$ and $Z$. For any point $y\in Y$ the subspace $\nu (\mu^{-1}(y))$
is a symplectic
leaf in $Z$. It carries nondegenerate symplectic structure. The same
is true  in the other
direction. Take any point $z\in Z$, then the subspace $\mu
(\nu^{-1}(z))$ is a
symplectic leaf in $Y$. Actually, in this paper we don't need the
full machinery
of dual pairs. Only one simple fact will be of importance for us.

\begin{lemma}
Let the pair of mappings $(\mu ,\nu)$ (\ref{DP}) be a dual pair.
Suppose that the Poisson bracket on $Y$ is equal to zero at the
point $y$. Under these conditions the restriction of the symplectic
form $\Omega$ on $X$ to the subspace $\mu^{-1}(y)$ coincides
with the pull back of the symplectic form $\omega_{y}$ on the
symplectic leave $\nu(\mu^{-1}(y))$ along the projection $\nu$:

\begin{equation}
\O \mid_{\mu^{-1}(y)}=\nu^{*}\omega_{y}.  \label{Oo}
\end{equation}
\end{lemma}
This lemma relates the symplectic structure of the reduced phase
space with the symplectic structure of the global space $X$ which
is usually much simpler.

A particular example of the conditions of {\em Lemma 1} is provided
by
the Hamiltonian reduction over the origin of the momentum mapping.
Indeed, the Poisson structure of the space $\g^{*}$ is described by
Kirillov-Kostant-Sourieu bracket:

\begin{equation}
\{y^{a},y^{b}\}=f^{ab}_{c}y^{c},     \label{KK}
\end{equation}
where $f^{ab}_{c}$ are structure constants of the Lie algebra $\g$.
At the
origin of  $\g^{*}$ coordinates $y^{c}$ are equal to zero and the
Poisson bracket
is obviously equal to zero for  any functions  on $\g^{*}$. It means
that {\em Lemma 1}
is applicable for  the moduli space of flat connections on a Riemann
surface
with marked points. The symplectic structure in question may be
investigated
using the relatively simple symplectic form (\ref{SAtot}) on the
space
${\cal A}_{g,n}^{tot}$. The subject of the next Section is how to
make this
description indeed efficient.

\section{Combinatorial description of the symplectic structure on the
moduli space}
\setcounter{equation}{0}
As it was pointed in subsection 2.3,  the pull-back of the canonical
symplectic
structure on the moduli space  to the space of flat connections on
the decorated
Riemann surface is easy to describe because it coincides with the
restriction
of the canonical symplectic structure on the space ${\cal
A}^{tot}_{g,n}$.
The drop back of this description is that we have to use flat
connections as
coordinates on the moduli space. The space of flat connections is
infinite
dimensional , whereas the moduli space is finite dimensional for
finite $g$
and $n$. So, we should look for more efficient coordinate mappings.
The simplest
example of such a mapping may be constructed in the following way.
Let us choose
a point $P$ on the Riemann surface which does not coincide with
marked points
$z_{i}$. One can define a subgroup of the gauge group
$G_{\Sigma}(P)$ by the
requirement:

\begin{equation}
G_{\Sigma}(P)=\{g\in G_{\Sigma}, \quad g(P)=I\}.  \label{GS}
\end{equation}
The quotient space

\begin{equation}
\m_{g,n}(P)=\Im_{g,n}/G_{\Sigma}(P)
\end{equation}
is already finite dimensional and admits efficient parametrization.

Let us draw a bunch of circles on the Riemann surface so that  there
is only one intersection  point $P$. In this bunch we have two
circles for each handle (corresponding to $a$- and $b$- cycles) and
one circle for each marked point. We shall denote the circles
corresponding to the $i$'s handle by $a_{i}$ and
$b_{i}$ ($i = 1, \dots ,g$) and we shall use symbols $m_{i}$ ($i =
1,\dots ,n$) for the circles surrounding marked points. We  assume
that the circles on $\Sigma$
are chosen in such a way that the only defining relation in
$\pi_{1}(\Sigma_{g,n})$ looks as

\begin{equation}
m_{1}\ldots m_{n} (a_{1}b_{1}^{-1}a_{1}^{-1}b_{1})\ldots
(a_{g}b_{g}^{-1}a_{g}^{-1}b_{g}) = id.
\label{abc}
\end{equation}

To each circle we assign the corresponding monodromy matrix
defined by the flat connection $A$.
Let us denote these matrices by $A_{i}$,$B_{i}$ and $M_{i}$ for $a$-,
$b$- and $m$-circles. The set of monodromy matrices provides
coordinates on $\m_{g,n}$ and  a
representation of the fundamental group $\pi_{1}(\Sigma_{g,n})$.
It implies the relation

\begin{equation}
M_{1}\ldots M_{n} (A_{1}B_{1}^{-1}A_{1}^{-1}B_{1})\ldots
(A_{g}B_{g}^{-1}A_{g}^{-1}B_{g})= I
\label{ABC}
\end{equation}
imposed on the values of $A_{i}$,$B_{i}$ and $M_{i}$.
Actually, monodromies $M_{i}$ are not arbitrary. They belong to
conjugancy
classes $C_{i}(G)$ defined by

\begin{equation}
M_{i}=u_{i}^{-1}C_{i}u_{i},  \label{MuCu}
\end{equation}
where

\begin{equation}
C_{i}=\exp{(\frac{2\pi}{k}D_{i})}.    \label{Ci}
\end{equation}

So the space $\m_{g,n}(P)$ is a subspace in
\begin{equation}
{\cal F}_{g,n}=G^{2g}\times \prod_{i=1}^{n}{C_{i}(G)} \label{calF}
\end{equation}
defined by the relation (\ref{ABC}).

The original moduli space may be represented as a quotient of
$\m_{g,n}$
over the residual gauge group which is isomorphic to the group $G$:

\begin{equation}
\m_{g,n}=\m_{g,n}(P)/G.
\end{equation}

It is convenient to define some additional coordinates $K_{i}$ on
${\cal F}_{g,n}$:

\begin{eqnarray}
K_{0}=I,\nonumber \\
K_{i}=M_{1}\dots M_{i},  1\leq i\leq n \nonumber \\
K_{n+2i-1}=K_{n+2i-2}A_{i}, \\
K_{n+2i}=K_{n+2i-1}B_{i}^{-1}A_{i}^{-1}B_{i}.\nonumber \label{KKK}
\end{eqnarray}
It follows from the equation (\ref{ABC}) that

\begin{equation}
K_{n+2g}=K_{0}=I.  \label{KKI}
\end{equation}

Unfortunately,  coordinates $A,B,M$ and $K$ are not  sufficient for
analysis of the symplectic form on the moduli space and we have to
introduce
a new space $\tilde{\F}$:

\begin{equation}
\tilde{\F}=G^{n+2g}\times H^{n+g}.  \label{Ft}
\end{equation}
Here $H$ is a Cartan subgroup of $G$. $\tilde{\F}$ may be
parametrized by
matrices $u_{i}, i=1,\dots ,n+2g$ from the group $G$ and by Cartan
elements
 $C_{i}, i=1,\dots ,n+g$. We define a projection from $\tilde{\F}$ to
$\F$ by
the formulae:

\begin{eqnarray}
M_{i}=u_{i}^{-1}C_{i}u_{i}, \nonumber\\
A_{i}=u_{n+2i-1}^{-1}C_{n+i}u_{n+2i-1}, \\
B_{i}=u_{n+2i}u_{n+2i-1}^{-1}. \nonumber  \label{Cuu}
\end{eqnarray}
 Let us call  $\tilde{\m}_{g,n}(P)$ the preimage of  $\m_{g,n}(P)$ in
$\tilde{\F}$.

After this lengthy preparations we are ready to formulate the main
result of this Section.

\begin{theorem}
The pull-back of the canonical symplectic form on $\m_{g,n}$ to
$\tilde{\m}_{g,n}(P)$
coincides with the restriction   of the following two-form defined on
$\tilde{\F}$:

\begin{eqnarray}
\O_{\F}=\frac{k}{4\pi}Tr [\sum_{i=1}^{n+2g}{\d
u_{i}u_{i}^{-1}C_{i}\wedge
\d u_{i}u_{i}^{-1}C_{i}^{-1}}-
\sum_{i=1}^{n+2g}{\d K_{i}K_{i}^{-1} \wedge \d K_{i-1}K_{i-1}^{-1}}+
\nonumber \\
+\sum_{i=1}^{g}{\d C_{n+i}C_{n+i}^{-1}\wedge (\d u_{n+2i}
u_{n+2i}^{-1}-\d u_{n+2i-1}u_{n+2i-1}^{-1})}]. \label{th1}
\end{eqnarray}
\end{theorem}
The rest of the Section is devoted to proof of {\em Theorem 1}.

{\em Proof}.

Let us cut the surface along every circle  $a_{i}, b_{i}, m_{i}$. We
get $n+1$ disconnected parts.
The first $n$  are similar. Each of them is a  neighborhood of the
marked point with  the cycle $m_{i}$ as a boundary. We denote these
disjoint parts by $P_{i}$. The last one is a polygon. There is no
marked points inside and the boundary is composed of $a$-,$b$-, and
$m$-cycles as it is prescribed by formula (\ref{abc}). We denote the
polygon by $P_{0}$.

Being restricted to $P_{0}$ a  flat connection $A$ becomes trivial:

\begin{equation}
A\mid _{P_{0}}= g_{0}^{-1}dg_{0}.   \label{gdg}
\end{equation}
For any other part $P_{i}$ we get a bit more complicated expression:

\begin{equation}
A\mid _{P_{i}}= \frac{1}{k}g_{i}^{-1}D_{i}g_{i}d\phi_{i}
+g_{i}^{-1}dg_{i}.   \label{gDg}
\end{equation}
We remind that $D_{i}$ is a diagonal matrix which characterizes the
orbit
attached to the marked point $z_{i}$. There is a set of consistency
conditions
which tells that the connection described by formulae
(\ref{gdg},\ref{gDg})
is actually smooth on the Riemann surface everywhere except the
marked points.
It means that when one approaches the cuts from two sides, one always
gets the
same value of $A$. To be explicit, let us consider the $m$-cycle
which surrounds
the marked point $z_{i}$. Comparison of equations
(\ref{gdg},\ref{gDg}) gives:

\begin{equation}
g_{0}^{-1}dg_{0}\mid_{m_{i}}=(\frac{1}{k}
g_{i}^{-1}D_{i}g_{i}d\phi_{i} +g_{i}^{-1}dg_{i})\mid_{m_{i}}.
\label{Con1}
\end{equation}
This equation may be easily solved:

\begin{equation}
g_{0}\mid_{m_{i}}=NMg_{i}\mid_{m_{i}},  \label{Conm}
\end{equation}
where $N$ is an arbitrary constant matrix and $M$ is equal to

\begin{equation}
M(\phi_{i})=\exp{(\frac{1}{k}D_{i}\phi_{i})}.  \label{Mphi}
\end{equation}
Now we turn to consistency conditions which arise when one considers
$a$- or $b$-cycles. In this case both sides of the cut belong to the
polygon $P_{0}$.
Let us denote the restrictions of  $g_{0}$ on the cut sides by
$g^{'}$ and $g^{''}$.
So we have:

\begin{equation}
g^{'-1}dg^{'}=g^{''-1}dg^{''}. \label{Con2}
\end{equation}
We conclude that the matrices $g^{'}$ and $g^{''}$ may differ only by
a constant left multiplier:

\begin{equation}
g^{''}=Ng^{'}.  \label{Conab}
\end{equation}

By now we considered connection $A$ in the region of the surface
where it is flat.
However,  it is not true at the marked points.  We calculate the
curvature in the region $P_{i}$ and  get a $\d$-function singularity:
\begin{equation}
F(z)\mid _{P_{i}}=\frac{2\pi}{k}g_{i}^{-1}D_{i}g_{i}\d (z-z_{i}).
\label{FPi}
\end{equation}

Equations (\ref{FPi},\ref{MF},\ref{MF0}) imply that the value
$g_{i}(z_{i})$ coincides
with the matrix $v_{i}$:

\begin{equation}
g_{i}(z_{i})=v_{i}.         \label{gu}
\end{equation}
Let us remind that $v_{i}$ diagonalizes the matrix $T_{i}$ attached
to the marked point $z_{i}$
by definition of the decorated Riemann surface.

Now we are prepared to consider  the symplectic structure on the
space of flat connections. First, let us rewrite the definition
(\ref{SAtot}) in the following way:

\begin{equation}
\O^{tot}=\o_{0}+\sum_{i=1}^{n}{\o_{i}}, \label{OOO}
\end{equation}
where the summands correspond to different parts of the Riemann
surface:

\begin{eqnarray}
\o_{0}=\frac{k}{4\pi}Tr\int_{P_{0}}{\d A\wedge \d A}, \nonumber \\
\o_{i}=\frac{k}{4\pi}Tr\int_{P_{i}}{\d A\wedge \d A}+\varpi_{i}.
\label{OOi}
\end{eqnarray}
The next step must be to substitute (\ref{gdg},\ref{gDg}) into
formulae (\ref{OOi}).
The following lemma provides an appropriate technical tool for this
operation.

\begin{lemma}
Let  $A$ be a $\g$-valued connection defined in the region $P$ of the
Riemann surface
$\Sigma$. Suppose that

\begin{equation}
A=g^{-1}Bg+g^{-1}dg.   \label{AB}
\end{equation}
Then the canonical symplectic form
\begin{equation}
\o_{P}=Tr\int_{P}{\d A\wedge \d A}  \label{OP}
\end{equation}
may be rewritten as

\begin{equation}
\o_{P}=Tr\int_{P}\{ \d B\wedge \d B+2\d [F_{B}\d gg^{-1}]\}
+Tr\int_{\partial P}\{ \d gg^{-1} d(\d gg^{-1})-\d [B\d gg^{-1}]\},
\label{lem2}
\end{equation}
where $F_{B}$ is a curvature of the connection $B$

\begin{equation}
F_{B}=dB-B^{2}. \label{FB}
\end{equation}
\end{lemma}
One can prove  {\em Lemma 2}  by straightforward calculation.

Let us apply {\em Lemma 2} to the polygon $P_{0}$. In this case $B=0$
and the answer reduces to

\begin{equation}
\o_{0}=\frac{k}{4\pi}Tr\int_{\partial P_{0}}\d g_{0}g_{0}^{-1} d(\d
g_{0}g_{0}^{-1}).  \label{O0}
\end{equation}

The boundary of the polygon $\partial P_{0}$ consists of $n+4g$
cycles (\ref{abc}). So actually
we have $n+4g$ contour integrals in the r.h.s. of  (\ref{O0}).

Now we use formula (\ref{lem2}) to rewrite symplectic structures
$\o_{i}$:

\begin{eqnarray}
\o_{i}=\frac{k}{4\pi}Tr\int_{\partial P_{I}}\{\d g_{i}g_{i}^{-1} d(\d
g_{i}g_{i}^{-1})-
\frac{2\pi}{k}\d [D_{i}\d g_{i}g_{i}^{-1}]\}-
\nonumber \\
-Tr\int_{P_{I}}\d\{D_{i}\d g_{i}g_{i}^{-1}\}\d (z-z_{i})+Tr D_{i}(\d
v_{i}v_{i}^{-1})^{2}.  \label{Oi}
\end{eqnarray}
The last term in (\ref{Oi}) represents Kirillov form attached to the
marked point $z_{i}$. Taking
into account relation (\ref{gu}) we discover that this term together
with the third term in (\ref{Oi})
cancel each other.

At this point it is convenient to denote the values of $g_{0}$
at the corners of the polygon. We enumerate the corners by the index
$i=0,\dots ,n+4g-1$
so that the end-points of the cycle $m_{i}$ are labeled by $i-1$ and
$i$. One can easily read
from formula (\ref{abc}) the enumeration of the ends of $a$- and
$b$-cycles
(see Fig. 1). For example,
the end-points of $a_{i}$ are labeled by $n+4(i-1)$ and $n+4(i-1)+1$,
whereas the end-points
of $a_{i}^{-1}$ entering in the same word are labeled by $n+4(i-1)+2$
and $n+4(i-1)+3$.
We denote the value of $g_{0}$ at the $i$'s corner by $h_{i}$.

Monodromies $A_{i}$,$B_{i}$ and $M_{i}$ may be expressed in terms of
$h_{i}$ as

\begin{eqnarray}
M_{i}=h_{i-1}^{-1}h_{i}, \\ \label{Mh}
A_{i}=h_{n+4(i-1)}^{-1}h_{n+4(i-1)+1}=
h_{n+4(i-1)+3}^{-1}h_{4(i-1)+2}, \\ \label{Ah}
B_{i}=h_{n+4(i-1)+1}^{-1}h_{n+4(i-1)+2}=h_{n+4i}^{-1}h_{4(i-1)+3}.
\label{Bh}
\end{eqnarray}

Let us remark that without loss of generality we can choose $g_{0}$
in such a way
that its value $h_{0}$ is equal to unit element in $G$.
After that  some of the corner values $h_{i}$ may be identified with
$K_{i}$;

\begin{equation}
K_{i}=\cases{
h_{i} & for $1\leq i\leq n$ \cr
h_{2i-n-1} & for $(i-n)$ odd \cr
h_{2i-n}& for $(i-n)$ even
}       \label{Kh}
\end{equation}

Our strategy is to adjust notations to the description of Poisson-Lie
symplectic
forms (see subsection 2.2).  Using formula (\ref{Conm}) one can
diagonalize $M_{i}$

\begin{equation}
M_{i}=u_{i}^{-1}C_{i}u_{i}.    \label{uCu}
\end{equation}
Here $u_{i}$ is the value of the variable $g_{i}$ at the point $P$.

Let us rewrite formula  (\ref{O0}) in the following way:

\begin{equation}
\o_{0}=\sum_{i=1}^{n}\varphi_{i}+\sum_{i=1}^{g}\psi_{i}. \label{O02}
\end{equation}

Here $\varphi_{i}$ is a  contribution corresponding to the marked
point:
\begin{equation}
\varphi_{i}=\frac{k}{4\pi}Tr\int_{m_{i}}\d g_{0}g_{0}^{-1} d(\d
g_{0}g_{0}^{-1}),  \label{varphi}
\end{equation}

and $\psi_{i}$ is a contribution of the handle:

\begin{equation}
\psi_{i}=\frac{k}{4\pi}Tr\int_{a_{i}b_{i}^{-1}a_{i}^{-1}b_{i}}\d
g_{0}g_{0}^{-1} d(\d g_{0}g_{0}^{-1}). \label{psi}
\end{equation}

First, we are going to evaluate the total contribution of the given
$M$-cycle
which is equal to a sum of two terms:

\begin{equation}
\O_{i}=\o_{i}+\varphi_{i}.  \label{Oop}
\end{equation}
Actually, each summand in (\ref{Oop}) includes an integral over the
$m$-cycle.
However, this sum of integrals is an integral of exact form and it
depends only
on some finite number of boundary values. This situation is typical
and will
repeat when we consider a contribution of a handle.

\begin{lemma}
The form $\o_{i}$ depends only on finite number of parameters and may
be written as

\begin{equation}
\o_{i}=\frac{k}{4\pi}Tr[C_{i}\d u_{i}u_{i}^{-1}\wedge C_{i}^{-1}\d
u_{i}u_{i}^{-1}-
\d K_{i}K_{i}^{-1}\wedge \d K_{i-1}K_{i-1}^{-1}].  \label{oi}
\end{equation}
\end{lemma}
To prove  {\em Lemma 3} one should substitute formula (\ref{Conm})
into
expression for $\varphi_{i}$, integrate by parts and compare the
result with  the expression for
$\o_{i}$. The integrals in  $\varphi_{i}$ and  $\o_{i}$ cancel each
other and after
rearrangements the boundary terms reproduce formula (\ref{oi}).

Now we turn to the contribution of  a handle $\psi_{i}$
into the symplectic form on the moduli space.
One can see that each $a$-cycle and each $b$-cycle enter twice into
expression
(\ref{O02}). These two contributions correspond to two sides of the
cut. As usual, the
result simplifies if we combine the contributions of two cut sides
together.

\begin{lemma}
Let  $g^{'}, g^{''}$ be two mappings from the segment $[x_{1},
x_{2}]$ into the group G
with boundary values $g^{'}_{1,2}, g^{''}_{1,2}$. Suppose that these
mappings  differ by the $x$-independent left multiplier

\begin{equation}
g^{''}=Ng^{'}.    \label{gNg}
\end{equation}
Then the following equality holds:

\begin{eqnarray}
\O_{\ [x_{1}, x_{2}]}=Tr\int_{x_{1}}^{x_{2}}{\d g^{''}g^{''-1} d(\d
g^{''}g^{''-1})}-
Tr\int_{x_{1}}^{x_{2}}{\d g^{'}g^{'-1} d(\d g^{'}g^{'-1})}= \nonumber
\\
=Tr(g_{1}^{'-1}\d g_{1}^{'}\wedge g_{1}^{''-1}\d
g_{1}^{''}-g_{2}^{'-1}\d g_{2}^{'}
\wedge g_{2}^{''-1} \d g_{2}^{''}).
\end{eqnarray}
\end{lemma}
Proof is straightforward.

Let us parametrize $A_{i}$ and $B_{i}$ as in (\ref{Cuu}):

\begin{equation}
A_{i}=u_{n+2i-1}^{-1}C_{n+i}u_{n+2i-1},\quad
u_{n+2i}=B_{i}u_{n+2i-1}. \label{uu}
\end{equation}

One of the motivations for such notations is the following identity:
\begin{equation}
B_{i}^{-1}A_{i}^{-1}B_{i}=u_{n+2i}^{-1}C_{n+i}^{-1}u_{n+2i}.
\label{BAB}
\end{equation}

In principle, one can introduce the following uniformal variables

\begin{eqnarray}
M_{n+2i-1}=A_{i}=u_{n+2i-1}^{-1}C_{n+i}u_{n+2i-1},\nonumber \\
M_{n+2i}=B_{i}^{-1}A_{i}^{-1}B_{i}=u_{n+2i}^{-1}C_{n+i}^{-1}u_{n+2i}.
\label{MM}
\end{eqnarray}
so that the defining relation (\ref{ABC}) looks as

\begin{equation}\label{MMM}
M_{1}\dots M_{n}M_{n+1}\dots M_{n+2g}=I.
\end{equation}
In these variables we treat handles and marked points in the same
way. Roughly
speaking, one handle produces two marked points which have the
inverse
values of $C$: $C_{1}=C_{n+i}$, $C_{2}=C_{n+i}^{-1}$. It resembles
the relation between the
double $D_{+}$ and  two orbits
of dressing transformations (see subsection 2.2). Using the
definition of $M$ (\ref{MM})
we can clarify the definition of $K_{i}$:

\begin{equation}
K_{i}=M_{1}\dots M_{i}. \label{KM}
\end{equation}

Now we turn to the contribution $\psi_{i}$ of a handle into
symplectic form
(\ref{O02}).

\begin{lemma}
The handle contribution into symplectic form depends only on the
values
of $g_{0}$ at the end-points of the corresponding $a$- and $b$-cycles
and
may be written as
\begin{eqnarray} \label{Psi}
\psi_{i}=\frac{k}{4\pi}Tr[C_{n+i}\d u_{n+2i-1}u_{n+2i-1}^{-1}\wedge
C_{n+i}^{-1}\d u_{n+2i-1}u_{n+2i-1}^{-1}- \nonumber \\ -\d
K_{n+2i-1}K^{-1}_{n+2i-1}\wedge \d
K_{n+2(i-1)}K^{-1}_{n+2(i-1)}+\nonumber \\
+C_{n+i}^{-1}\d u_{n+2i}u_{n+2i}^{-1}\wedge C_{n+i}\d
u_{n+2i}u_{n+2i}^{-1}
-\d K_{n+2i}K^{-1}_{n+2i}\wedge \d K_{n+2i-1}K^{-1}_{n+2i-1}+ \\
+\d C_{n+i}C_{n+i}^{-1}\wedge (\d u_{n+2i-1}u_{n+2i-1}^{-1}-\d
u_{n+2i} u_{n+2i}^{-1})].  \nonumber
\end{eqnarray}
\end{lemma}
If we take into account {\em Lemma 4}, the proof of {\em Lemma 5}
becomes straightforward
but long calculation. Let us remark that the terrible formula
(\ref{Psi}) contains two copies
of the marked point contribution (\ref{oi}) with parameters $C_{n+i}$
and $C_{n+i}^{-1}$.  The last term includes $\d C_{n+i}C_{n+i}^{-1}$
and coincides with the corresponding additional term in formula
(\ref{FD}) for the symplectic form on the double $D_{+}$.

Summarizing {\em Lemma 3} and {\em Lemma 5}  we get the proof of {\em
Theorem 1}
completed.

\section{Equivalence to Poisson-Lie symplectic structure}
\setcounter{equation}{0}

Formula (\ref{th1}) contains  cross-terms  with different indices
$i$. In this Section we represent the canonical symplectic structure
as a direct sum of several terms. Using subsection 2.2, each term may
be identified with either Kirillov form for the Poisson-Lie group
$G^{*}$ or symplectic form on the Heisenberg double $D_{+}$ of the
Poisson-Lie group $G$.  To achieve this result we have to make a
change of variables. The new set of variables is designed to
''decouple''
contributions of different handles and marked points.

The following remark is important for understanding of the
construction of
decoupled variables. Monodromy matrices $M_{i}$, $A_{i}$ and $B_{i}$
are
elements of the group $G$. In accordance with this fact we use
$G$-multiplication
to define the variables $K_{i}$ (\ref{KM}) and to constraint
monodromies (\ref{ABC}).
On the other hand, natural variables for description of orbits of
dressing transformations
or double $D_{+}$ must belong to $G^{*}$. In Section 2 we defined the
mapping
$\alpha: G^{*}\rightarrow G$. Unfortunately, $\alpha$ is not a group
homomorphism.
So, we would face difficulties applying $\alpha$ to identities
(\ref{KM},\ref{ABC}).
This is a motivation to introduce a notion of a weak group
homomorphism.

\begin{definition}
Let $G$ and $G^{'}$ be two groups. A set of mappings

\begin{equation}
\alpha^{(n)}: G^{n}\rightarrow  G^{'n} \label{GG}
\end{equation}
is called a weak  homomorphism if the following diagram is
commutative for any $i$:

\begin{eqnarray}
G^{n}\quad\ \ \stackrel{\alpha^{(n)}}{\longrightarrow}\ \ \quad
G^{'n}\ \quad
\nonumber\\
{\scriptstyle  \bf m}_{i}\downarrow \ \ \qquad\qquad {\scriptstyle
\bf m}^{'}_{i}
\downarrow\qquad \\
G^{n-1} \ \ \stackrel{\alpha^{(n-1)}}{\longrightarrow}\quad G^{'n-1}\
 .\nonumber
\end{eqnarray}
Here ${\bf m}_{i}$ and  ${\bf m}^{'}_{i}$ are multiplication mappings
in
$G$ and $G^{'}$ correspondingly which map the product of $n$ copies
of the group
into the product of $n-1$ copies:

\begin{eqnarray}
{\bf m}_{i}: (g_{1},\dots ,g_{i}, g_{i+1},\dots ,g_{n})\rightarrow
(g_{1},\dots ,g_{i}g_{i+1},\dots ,g_{n}):\nonumber \\
{\bf m}^{'}_{i}: (g^{'}_{1},\dots ,g^{'}_{i}, g^{'}_{i+1},\dots
,g^{'}_{n})\rightarrow (g^{'}_{1},\dots ,g^{'}_{i}g^{'}_{i+1},\dots
,g^{'}_{n}):\nonumber \\
\end{eqnarray}
\end{definition}

The mapping $\alpha$ (\ref{LL}) may be considered as a first mapping
of a weak
homomorphism from $G^{*}$ to $G$. To define the other mappings
$\alpha^{(n)}$ we
introduce  the products

\begin{equation}
K_{\pm}(i)=L_{\pm}(1)\dots L_{\pm}(i).   \label{KLL}
\end{equation}
The action of $\alpha^{(n)}$ looks as follows.
A tuple  $(L_{+}(i), L_{-}(i))\in G^{*}, i=1,\dots n$ is
mapped  into the tuple $M_{i}\in G, i=1,\dots n$:

\begin{equation}
M_{i}=K_{-}(i-1)L_{i}K_{-}(i-1)^{-1}. \label{KLK}
\end{equation}
Here $L_{i}$ is the image of the pair $(L_{+}(i), L_{-}(i))$ under
the action of $\alpha$:

\begin{equation}
L_{i}=L_{+}(i)L_{-}(i)^{-1}.  \label{LLL}
\end{equation}
One can easily check that the set of mappings (\ref{KLK}) satisfies
the requirements of
a weak homomorphism.

The next step is to implement the definition (\ref{KLK}) to the space
 $\tilde{\F}$.
Let us introduce a set of variables on $\tilde{\F}$ which consists of
$v_{i}, i=1,\dots ,n+2g$
taking values in $G$ and  $C^{'}_{i}, i=1,\dots ,n+g$ taking values
in $H$.
In addition we introduce the elements of $G^{*}$:

\begin{eqnarray} \label{LLLL}
L_{i}=v_{i}^{-1}C^{'}_{i}v_{i}\quad  for 1\leq i\leq n;\nonumber \\
L_{n+2i-1}=v_{n+2i-1}C^{'}_{n+i}v_{n+2i-1}\quad for1\leq i\leq g;  \\
L_{n+2i}=v_{n+2i}C^{'-1}_{n+i}v_{n+2i}\quad for 1\leq i\leq g.
\nonumber
\end{eqnarray}
together with their Gauss components  (\ref{LL}). So, we have natural
variables to
describe $n$ copies of the orbit of dressing transformations in
$G^{*}$ and $g$ copies of the Heisenberg double. The canonical
symplectic form on this object is equal to
the sum of symplectic forms for each copy of the orbit (\ref{FG}) and
each
copy of double (\ref{FD}):

\begin{equation}
\O_{PL}=\sum_{i=1}^{n}{\vartheta(u_{i},C^{'}_{i})}+
\sum_{i=1}^{g}{\theta(u_{n+2i-1},u_{n+2i},
C^{'}_{n+i})}. \label{OPL}
\end{equation}
Let us compare the forms (\ref{th1}) and  (\ref{OPL}). Motivated by
the definition
(\ref{KLK}) we introduce the mapping $\sigma: \tilde{\F}\rightarrow
\tilde{\F}$ defined
by the relations:

\begin{eqnarray}
u_{i}=v_{i}K^{-1}_{-}(i-1),\quad
C_{i}=C^{'}_{i}.  \label{Sig}
\end{eqnarray}
Here $K_{-}(i)$ are defined as in (\ref{KLL}). It is easy to see that
 the mapping $\sigma$
induces the mapping $\alpha^{(n+2g)}$  from the set of pairs
$(L_{+}(i),  L_{+}(i))$ into the
set of monodromies $M_{i}$.  It is guaranteed by the definition of
weak homomorphism that $G$-product in the relation (\ref{ABC}) is now
replaced by $G^{*}$-product:

\begin{equation}
K_{\pm}(n+2g)=L_{\pm}(1)\dots L_{\pm}(n+2g)=I.  \label{Kpm}
\end{equation}
Equation (\ref{Kpm}) defines the preimage of $\tilde{\m}_{g,n}$  in
$\tilde{\F}$
with respect to the mapping $\sigma$. It is worth mentioning that
the matrices $K_{i}$
from the previous Section may be represented as

\begin{equation}
K_{i}=K_{+}(i)K_{-}(i)^{-1}.  \label{Ki}
\end{equation}
This also a consequence of the definition of weak homomorphism.
Indeed, $K_{i}$  has
been defined as a product in $G$ of the first $i$ monodromies.
Formula (\ref{KLL})
defines a product in $G^{*}$ of $i$ first elements   $(L_{+}(i),
L_{+}(i))$. Using the basic
property of weak homomorphism $(i-1)$ times we check (\ref{Ki}).

The mapping $\sigma$ provides a possibility to compare two-forms
$\O_{\F}$ and
$\O_{PL}$.

\begin{lemma}
The two-forms $\O_{\F}$ is  propotional to the pull-back of the form
$\O_{PL}$
along the mapping $\sigma$:
\begin{equation}
\O_{F}=\f  \sigma^{*}(\O_{PL}). \label{lem7}
\end{equation}
\end{lemma}

{\em Lemma 6} may be proved by straightforward calculation. {\em
Theorem 1} and
{\em Lemma 6} imply the following theorem which is the main result of
this paper.

\begin{theorem}
Being restricted to the subset  (\ref{Kpm}), the direct sum of $n$
copies of Kirillov
symplectic form on the orbit of dressing transformations in $G^{*}$
and $g$ copies of the canonical form on the Heisenberg double of the
group $G$ coincides up to a scalar
multiplier with the pull-back of the canonical symplectic form on the
moduli space of flat connections on the Riemann surface of genus $g$
with $n$ marked points.
\end{theorem}

\section{Conclusions}
As we promised in Introduction, the symplectic form on the moduli
space of flat connections
may be split into $n$ pieces corresponding to the orbits of dressing
transformations and $g$
pieces corresponding to the copies of the Heisenberg double. By the
principle of orbit-representation correspondence \cite{Kir} one
should assign some irreducible
representations $I_{i}$ of the quantum group $U_{q}(\g)$ to the
orbits and the regular
representation $\Re$ to each copy of the Heisenberg double. Taking
into account
the constrain (\ref{Kpm}) which means that the representation of the
total spin is trivial,
we have a complete quasi-classical analogue of formula (\ref{I1}).

\section*{Acknowledgements}
This work has been initiated during Summer School on Gravitation and
Quantization, Les Houches 1992. A.A. is especially grateful to
G.Zuckermann
for stimulating discussions. We would like to thank L.D.Faddeev,
K.Gawedzki,
V.Fock, M.A.Semenov-Tian-Shansky and A.Weinstein for interest to this
work. Special thanks to A.J.Niemi for perfect conditions in Uppsala.

\begin{picture}(400,600)(-200,-300)

\put(0,160){\line(4,-1){28}}
\put(52,147){\line(4,-1){28}}
\put(80,140){\vector(1,-1){60}}
\put(140,80){\vector(1,-4){20}}
\put(140,-80){\vector(1,4){20}}
\put(80,-140){\vector(1,1){60}}
\put(80,-140){\vector(-4,-1){80}}
\put(0,-160){\vector(-4,1){80}}
\put(-80,-140){\line(-1,1){20}}
\put(-120,-100){\line(-1,1){20}}
\put(-140,-80){\vector(-1,4){20}}
\put(-160,0){\vector(1,4){20}}
\put(-140,80){\line(1,1){20}}
\put(-100,120){\line(1,1){20}}
\put(-80,140){\vector(4,1){80}}

\put(-170,38){$M_{1}$}
\put(-50,157){$M_{i}$}
\put(110,118){$B_{j-1}$}
\put(155,38){$A_{j}$}
\put(155,-38){$B_{j}^{-1}$}
\put(110,-122){$A_{j}^{-1}$}
\put(35,-165){$B_{j}$}
\put(-50,-165){$A_{j+1}$}
\put(-170,-42){$B_{g}$}

\put(-155,-3){$h_{0}=h_{n+4g}$}
\put(-137,72){$h_{1}$}
\put(-80,130){$h_{i-1}$}
\put(-4,147){$h_{i}$}
\put(23,133){$h_{n+4(j-1)-1}$}
\put(93,73){$h_{n+4(j-1)}$}
\put(100,-2){$h_{n+4(j-1)+1}$}
\put(80,-78){$h_{n+4(j-1)+2}$}
\put(25,-135){$h_{n+4(j-1)+3}$}
\put(-10,-150){$h_{n+4j}$}
\put(-77,-136){$h_{n+4j+1}$}
\put(-135,-80){$h_{n+4g-1}$}

\put(-226,-3){$K_{0}=K_{n+2g}$}
\put(-156,83){$K_{1}$}
\put(-105,144){$K_{i-1}$}
\put(-6,164){$K_{i}$}
\put(142,80){$K_{n+2(j-1)}$}
\put(162,-2){$K_{n+2(j-1)+1}$}
\put(-10,-172){$K_{n+2j}$}
\put(-100,-152){$K_{n+2j+1}$}

\put(-10,0){\Huge P}
\put(0,-5){\bf 0}

\end{picture}
\begin{center}
{\Large {\bf  Fig. 1}}
\end{center}

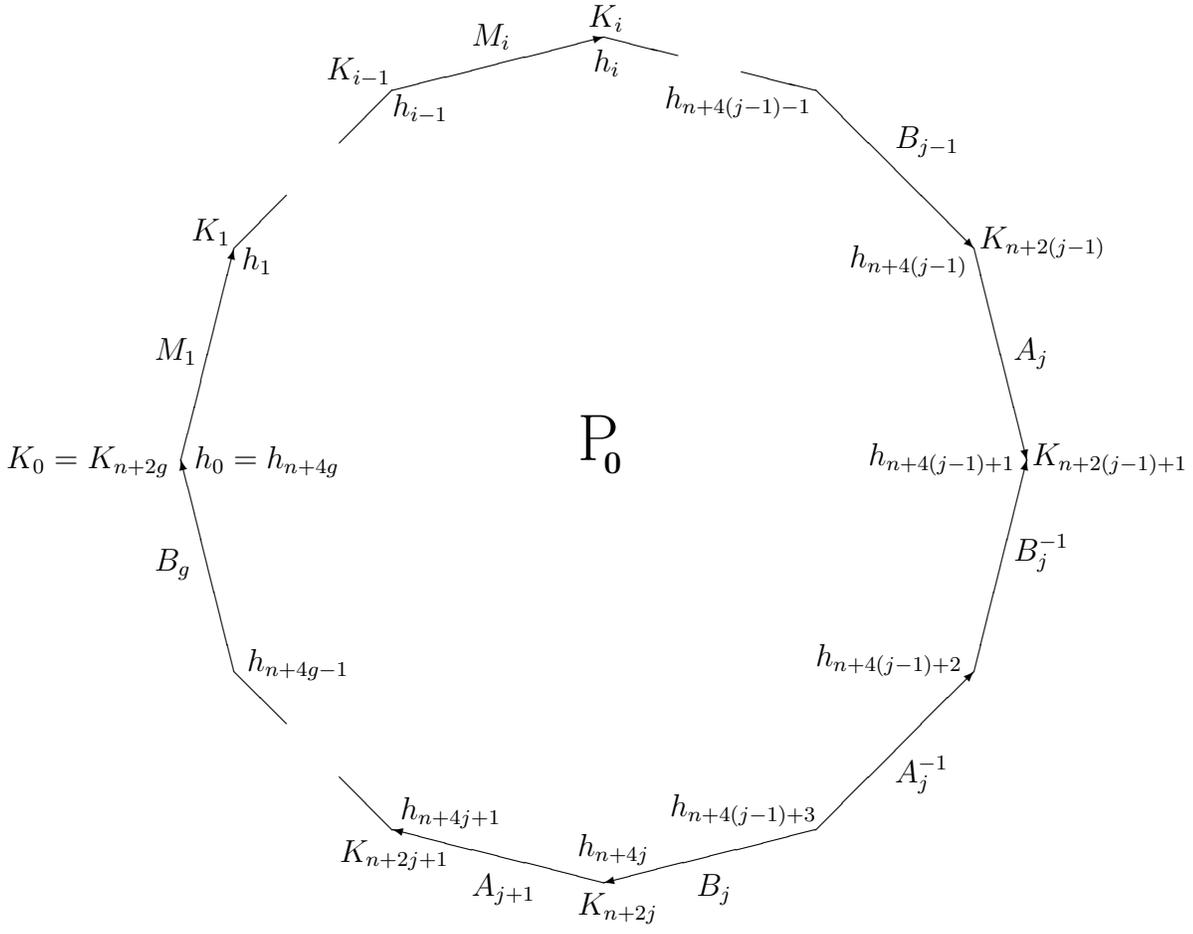
\end{document}